\documentclass[]{interact}

\usepackage{amsmath, amssymb, bm}
\let\chapter\section
\usepackage[ruled,vlined]{algorithm2e}
\usepackage[numbers,sort&compress]{natbib}
\bibpunct[, ]{[}{]}{,}{n}{,}{,}
\makeatletter
\def\NAT@def@citea{\def\@citea{\NAT@separator}}
\makeatother
\usepackage{epstopdf}

\begin{document}
\title{\bf Singhing with Confidence: Visualising the Performance of Confidence Structures}
\author{Alexander Wimbush\hspace{.2cm}\\
	a.wimbush@liverpool.ac.uk\\
	Institute for Risk and Uncertainty, University of Liverpool\\
	and\\
	Nicholas Gray\\
	Institute for Risk and Uncertainty, University of Liverpool\\
	and\\
	Scott Ferson\\
	Institute for Risk and Uncertainty, University of Liverpool}
\maketitle
\section*{Acknowledgements}
This work was supported by the EPSRC programme grant `Digital twins for improved dynamic design', EP/R006768/1, and the EPSRC and ESRC Centre for Doctoral Training in Quantification and Management of Risk and Uncertainty in Complex Systems and Environments, EP/L015927/1.

Word Count = 4815

\newpage
\begin{abstract}
Confidence intervals are an established means of portraying uncertainty about an inferred parameter and can be generated through the use of confidence distributions. For a confidence distribution to be ideal, it must maintain frequentist coverage of the true parameter. This can be represented for a precise distribution by adherence to a cumulative unit uniform distribution, referred to here as a Singh plot. This manuscript extends this to imprecise confidence structures with bounds around the uniform distribution, and describes how deviations convey information regarding the characteristics of confidence structures designed for inference and prediction. This quick visual representation, in a manner similar to ROC curves, aids the development of robust structures and methods that make use of confidence. A demonstration of the utility of Singh plots is provided with an assessment of the coverage of the ProUCL Chebyshev upper confidence limit estimator for the mean of an unknown distribution.
\end{abstract}

\begin{keywords}
Confidence, Coverage, Imprecise, Inference, Prediction
\end{keywords}

\section{Introduction}
Inferring the value of a model parameter or the output of a stochastic system is a common aspect of many scientific endeavours. In a frequentist view, this can be accomplished through the use of intervals which are guaranteed to bound the answer with a minimum given probability. These inferences and predictions can then be utilised with confidence, knowing that they will be correct with some minimum rate. Hence the name confidence intervals. Some strategies for generating these intervals include confidence distributions \cite{Schweder2002}, boxes and structures \cite{Balch2012} which can be used for computation, preserving the frequentist properties throughout \cite{ferson2014computing}.

However, this property only holds when the strategy for generating these intervals is valid. This could perhaps be proven mathematically, but in many cases such a proof may be excessively difficult to develop. If a person were to question whether the chosen strategy is valid, they may simply have to accept that they might not be able to verify this themselves. Or they may consider whether an alternative strategy is superior, but have no means of investigating this without significant effort. In these cases, there is a need for a means of validating that the strategy will produce intervals with this coverage property in a manner that is easy and interpretable.

This paper introduces the notion of confidence as an expression of the coverage property, and formalises the creation of plots to assess this. Various properties of these structures can be inspected, allowing new structures to be proposed, developed, and utilised for computation with confidence. For ease of interpretation, this paper will consider a case of inference about a distributional parameter $\theta_0$, though confidence intervals may also be drawn for predictions of the value of a sample $x_0$.

\section{Confidence Distributions}
\label{sec:ConfDists}
Confidence intervals are so named due to the fact that they provide assurance that a desired minimum proportion $\alpha$ of inferences about an uncertain parameter $\theta_0\in\Theta$ from a given length-$n$ dataset $\bm{x}=\{x_1,\dots,x_n\}\sim\text{F}(\theta_0)$ will be correct so long as the distributional assumptions hold. Generating these intervals is most commonly performed using a valid confidence distribution $\text{C}(\theta,\bm{x}):\Theta\to[0,1]$. For a desired confidence level $\alpha\in[0,1]$, an interval $\bm{\alpha}=\left[\underline{\alpha},\overline{\alpha}\right]\subseteq[0,1]$ is created such that $\overline{\alpha}-\underline{\alpha}=\alpha$. An $\alpha$-level confidence interval is then defined as
\begin{equation}\label{eq:ConfInv}
    \bm{\theta}=\left[\underline{\theta},\overline{\theta}\right]=\left[\text{C}^{-1}(\underline{\alpha},\bm{x}),\text{C}^{-1}\left(\overline{\alpha},\bm{x}\right)\right]
\end{equation}

These intervals are valuable for statistical inference and prediction. One such use case may be for estimation of the true mean value $\mu_0\in M$ of a normal distribution based on a sample of independent and identically distributed observations $\bm{x}=\{x_1,\dots,x_n\}\sim\text{N}(\mu_0, \sigma)$, where $\sigma$ and $\mu_0$ are unknown. While a point value estimate $\hat{\mu}_0=\bar{x}=\frac{1}{n}\sum{\bm{x}}$ is a useful statistic, the interpreter should have no confidence that future samples will share this precise mean since the target distribution is continuous and has non-zero variance. 

A lack of confidence here implies that the probability that this point estimate is the true mean is zero. Naturally the probability that future samples from this continuous distribution will have a mean bounded by a precise value is 0. To resolve this, an $\alpha$-level confidence interval generated using Equation \ref{eq:ConfInv} could be used as an estimate of $\mu_0$ with knowledge that at least an $\alpha$ proportion of such inferences will be correct. One means of generating such intervals is to draw them from the following confidence distribution\cite{singh2007confidence}:
\begin{equation}
    \text{C}(\mu,\bm{x})=\text{T}\left(\frac{\mu-\bar{x}}{s_{\bm{x}}/\sqrt{n}};n-1\right)
    \label{eq:NormPivot}
\end{equation}
where $\text{T}(\theta;n)$ is the cumulative probability density of a Student's t-distribution with $n$ degrees of freedom evaluated at $\theta$, and $s_{\bm{x}}$ is the standard deviation of the sample.
This then produces a cumulative distribution mapping the support to the unit interval $\text{C}(\mu,\bm{x}):M\to[0,1]$.
This function represents the confidence level $\alpha$ of a one-sided interval estimate $\bm{\mu}=[\min\left(M\right),\mu]$ of the parameter.

The $\alpha$-level confidence intervals that can be drawn from this distribution are not necessarily unique for a given $\alpha$. Generally a strategy for generating confidence intervals would include some limit on the $\bm{\alpha}$ intervals to ensure that each value of $\alpha$ corresponds to a unique interval on $M$. Some examples include one sided intervals $[0,\alpha]$  or $[1-\alpha,1]$, and centred intervals $[0.5-\frac{\alpha}{2}, 0.5+\frac{\alpha}{2}]$. With this defined, $\text{C}^{-1}(\bm{\alpha},\bm{x}):[0,1]\to M$ produces unique intervals with a desired confidence level $\alpha$.

\section{Validating Confidence Distributions}

It was asserted in Section  that Equation \ref{eq:NormPivot} is a confidence distribution, and that therefore the approach to generating confidence intervals defined above is valid. But this is not immediately apparent from Equation \ref{eq:NormPivot} alone. An analyst may wish to validate that the distribution they are using, or being instructed to use, is in fact a valid confidence distribution. Until a distribution is confirmed to maintain the property of coverage it should be considered a proposed distribution, denoted with an asterisk $\text{C}^*(\cdot)$. Verifying this property would allow its use as a confidence distribution for reliable inference. It should be noted that Equation \ref{eq:NormPivot} is a well known confidence distribution developed by Gosset over a century ago \cite{student1908probable}. The intent here is not to question the validity of this distribution, but to demonstrate the properties of a valid confidence distribution so that invalid distributions can be identified by comparison.

The first property required of a valid confidence distribution for inference about a parameter $\theta\in\Theta$ is that it must be a cumulative distribution across $\Theta$, and it should be apparent that Equation \ref{eq:NormPivot} satisfies this criteria. However, the second criteria required by Singh \cite{singh2007confidence} is not so simple to verify, and requires that at the true parameter value, for a given dataset $\bm{x}$, $\text{C}^*(\theta_0,\bm{x})$ follows the uniform distribution U$(0,1)$.
This effectively restricts a valid confidence distribution to the following condition:

\begin{equation}
    \text{Pro}(\text{C}(\bm{\theta},\bm{x}\sim\text{F}(\theta_0))\geq\alpha)=\alpha; \forall \bm{\theta}\ni\theta_0
\end{equation}

This prevents confidence structures which produce vacuous intervals from being considered valid. Confirming adherence to these properties allows true confidence intervals to be generated as defined in Section \ref{sec:ConfDists}. But this is not necessarily a simple task, and may indeed be prohibitively difficult mathematically. An alternative is to use a Monte Carlo approach to generate values of $\text{C}^*(\theta_0,\bm{x})$ and plotting the resulting values against the U$(0,1)$ distribution. This approach is here referred to as a Singh plot, in reference to the work of Professor Kesar Singh, a review of which was compiled by Babu\cite{JogeshBabu2014}.

\section{Singh Plots}

Singh plots can be used to calculate and visualise the performance of a proposed structure in terms of both coverage and aspects of conservatism.  
Once a proposed structure can be demonstrated to achieve coverage using the Singh plot, it can be used for reliably inferring confidence intervals about a parameter. Indications of over and under-confidence and the impact of sample sizes will also be apparent and may prompt the use or development of different structures. 

A Singh plot is generated by numerically estimating the minimum coverage probability of a series of $\alpha$ level confidence intervals. This is performed on a known distribution with defined parameters. Keeping with the proposed distribution for the mean $\mu_0$ of a normal distribution given in Equation \ref{eq:NormPivot}, since it is an unbounded continuous distribution if it can be demonstrated that the coverage is maintained when $\mu_0$ is known then the same should hold for cases where $\mu_0$ is uncertain.

A series of $m$ sample sets of length $n$ are drawn from the target normal distribution with defined parameters $\bm{X}=\{\bm{x}_1,\dots,\bm{x}_m\}, \bm{x}_i\sim\text{N}(\mu_0,\sigma)$. The proposed confidence distribution is then used to generate an interval with a defined strategy that has the minimum possible confidence whilst still covering the known true value. The one-sided strategy $\bm{\alpha}=[0,\overline{\alpha}=\text{C}^*(\mu_0,\bm{x})]$ provides such an interval for this distribution. Alternatively, an upper bounded interval could be used, and would have a very similar interpretation. This is due to the fact that the first derivative of the confidence distribution $\text{C}^{*'}(\mu,\bm{x})$ is at its minimum at the extremes of the support for the proposed distribution. If disjoint intervals are permitted the worst case interval is likewise one which extends to the extremes of the support but which rejects the central portion of the distribution.

This process is repeated to produce $m$ intervals with corresponding confidence values representing the minimum confidence required to bound the true value. Ordering these values and plotting them against a cumulative unit uniform distribution produces a plot of the minimum coverage probability for a given $\alpha$-level confidence interval. This is the Singh plot as described in Equation \ref{eq:SinghPlot} and Algorithm \ref{algo:SinghPlots}, which visually demonstrates whether criteria 2 is met. The interpretation is similar to that of a receiver operating characteristic curve, though the optimal case is represented by adherence to the diagonal rather than the top left of the plot. The receiver operating characteristic has utility in that is conveys a great deal of information visually in a manner that is easily interpretable \cite{hoo2017roc}. Singh plots allow a similar ease of interpretation when assessing confidence structures.

\begin{equation}\label{eq:SinghPlot}
    \text{S}(\bm{\alpha};\theta_0)=\text{Pro}\left(\text{C}^{*-1}(\bm{\alpha},\bm{x}\sim\text{F}(\theta_0))\ni\theta_0\right); \bm{x}\in\bm{X}
\end{equation}
\IncMargin{1em}
\begin{algorithm}
    \SetAlgoLined
    \SetKwInOut{Input}{input}\SetKwInOut{Output}{output}
    \Input{$C^*\>\>\leftarrow$Proposed confidence structure\\
        $\text{f}(\bm{\theta})\leftarrow$Target distribution taking parameters $\bm{\theta}$\\
        $\theta_0\>\>\>\leftarrow$True value of parameter of interest}
    \Output{Singh plot for visual assessment of confidence structure properties}
    \For{$i \in {1,\dots,m}$}{
        Generate sample: $\bm{x}={x_1, \dots, x_n}\sim\text{f}(\bm{\theta})$\;
        Calculate minimum required confidence for coverage: $s_i = \text{C}^*(\theta_0,\bm{x})$
    }
    Plot empirical CDF of $s$\\
    Plot CDF of $U(0,1)$ for comparison
    \caption{Generation of a Singh plot}\label{algo:SinghPlots}
\end{algorithm}\DecMargin{1em}

\begin{figure}[!ht]
    \centering
    \includegraphics[width=\linewidth]{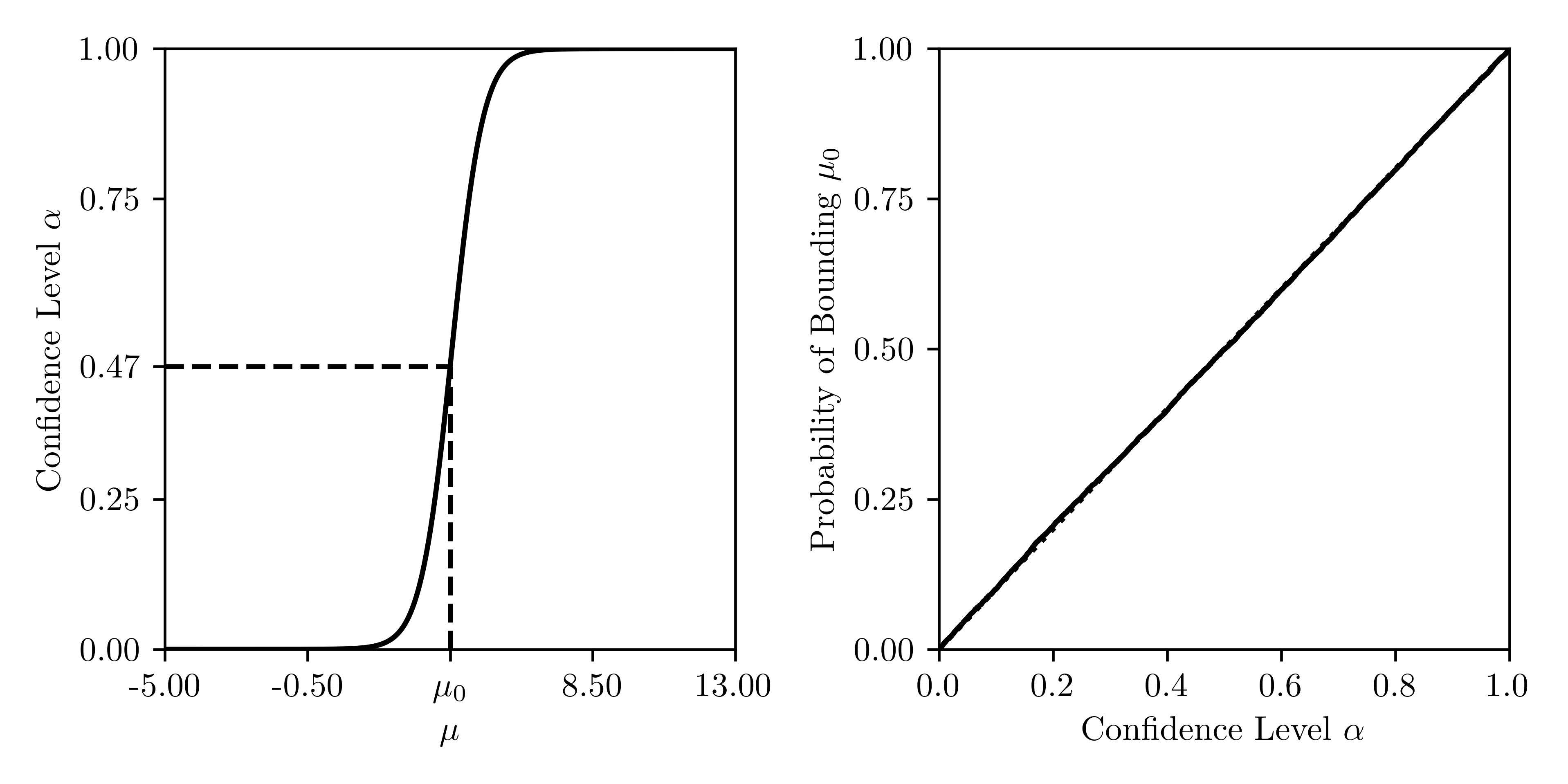}
    \caption{}
    \label{fig:NormCDist}
\end{figure}

For example, the confidence distribution in Equation \ref{eq:NormPivot} produces the Singh plot shown on the right side of Figure \ref{fig:NormCDist}. The left of the figure demonstrates $\text{C}^*(\mu,\bm{x})$ for one of the sample sets $\bm{x}$ for which the true value has been assigned a confidence value of $\text{C}^*(\mu_0,\bm{x})=0.65$. This indicates that a one-sided confidence interval would require a confidence level of at least $\alpha=0.65$ in order to bound the true value. The figure on the right extends this to $m=10^4$ samples of the same distribution, and indicates that $\text{Pro}(\text{C}^*(\bm{\mu},\bm{x})\geq\alpha)\approx\alpha; \forall \bm{\mu}\ni\mu_0$. There are deviations from the $\text{U}(0,1)$ distribution, but these are very slight and more likely to be dependent on $m$ than the proposed distribution. Note that since this is performed with a fixed mean and variance. Whilst it wouldn't be informative in this case to demonstrate coverage with variations in these parameters (since this structure is well established), this may be necessary depending on the application. The algorithm for this approach is provided in Section \ref{sec:Global}.

These Singh plots give a clear visual cue as to the reliability of these confidence structures. They allow any proposed structure to be quickly assessed to ensure that anyone looking to make use of them can do so from an informed perspective regardless of their mathematical capabilities. The result here is shown for demonstrative purposes, this confidence structure is known to perform at all confidence levels for one or two-sided intervals. The utility of the Singh plot becomes apparent when used to assess the properties of structures which may have unknown or poorly understood properties.

Producing a Singh plot according to Algorithm \ref{algo:SinghPlots} may be preferable to a conventional approach where confidence intervals are generated and then the rate at which they cover the true value is estimated. This is due to the fact that a Singh plot allows evaluation of coverage at all confidence levels rather than just one. It also allows visibility over whether or not a particular structure is conservative at some confidence levels and appropriately calibrated at others, or perhaps over-confident at some levels and conservative at others. This is particularly useful in the development of new structures which must maintain coverage. Singh plots are a quick and simple way to check whether the theory works in practice.

\subsection{Proposed Bernoulli Confidence Distribution Singh Plot}
A demonstration of a viable proposed distribution is shown in the preceding section, but how would an interpreter understand that the distribution is not viable? A simple example is that of inference about the rate parameter $\theta\in\Theta$ of a Bernoulli distribution. 
A sample drawn from a Bernoulli process using this distribution can be used to generate a Bayesian posterior from a conjugate Jeffreys prior. This would take the form a beta distribution with parameters $a=0.5$ and $b=0.5$. This produces the following proposed confidence distribution for inference about the true parameter $\theta_0$ given a sample set $\bm{x}=\{x_1,\dots,x_n\}, x_i\sim\text{Bin}(N=1, p=\theta_0)$ where $\text{Bin}(N=1, p)$ is a single observation binomial distribution with rate parameter $p$:
\begin{equation}
    \text{C}^*\left(\theta,\bm{x}\right)=\text{B}\left(\theta;a=\sum{\bm{x}}+0.5, b=n - \sum{\bm{x}} +0.5\right)
    \label{eq:BernPPropCDist}
\end{equation}
where B$(\theta;a, b)$ is the cumulative density of a beta distribution with parameters $a, b$ evaluated at $\theta$. Equation \ref{eq:BernPPropCDist} can be assessed as a proposed confidence distribution in a similar manner as the distribution in Equation \ref{eq:NormPivot}. A collection of $m$ Sample sets are drawn from the target distribution and used to generate one-sided $[0,\text{C}^*(\theta_0,\bm{x})]$ confidence intervals. Again, ordering and plotting the confidence levels of these intervals produces a Singh plot, which can be quickly checked to confirm that coverage is maintained.

\begin{figure}[!ht]
    \centering
    \includegraphics[width=\linewidth]{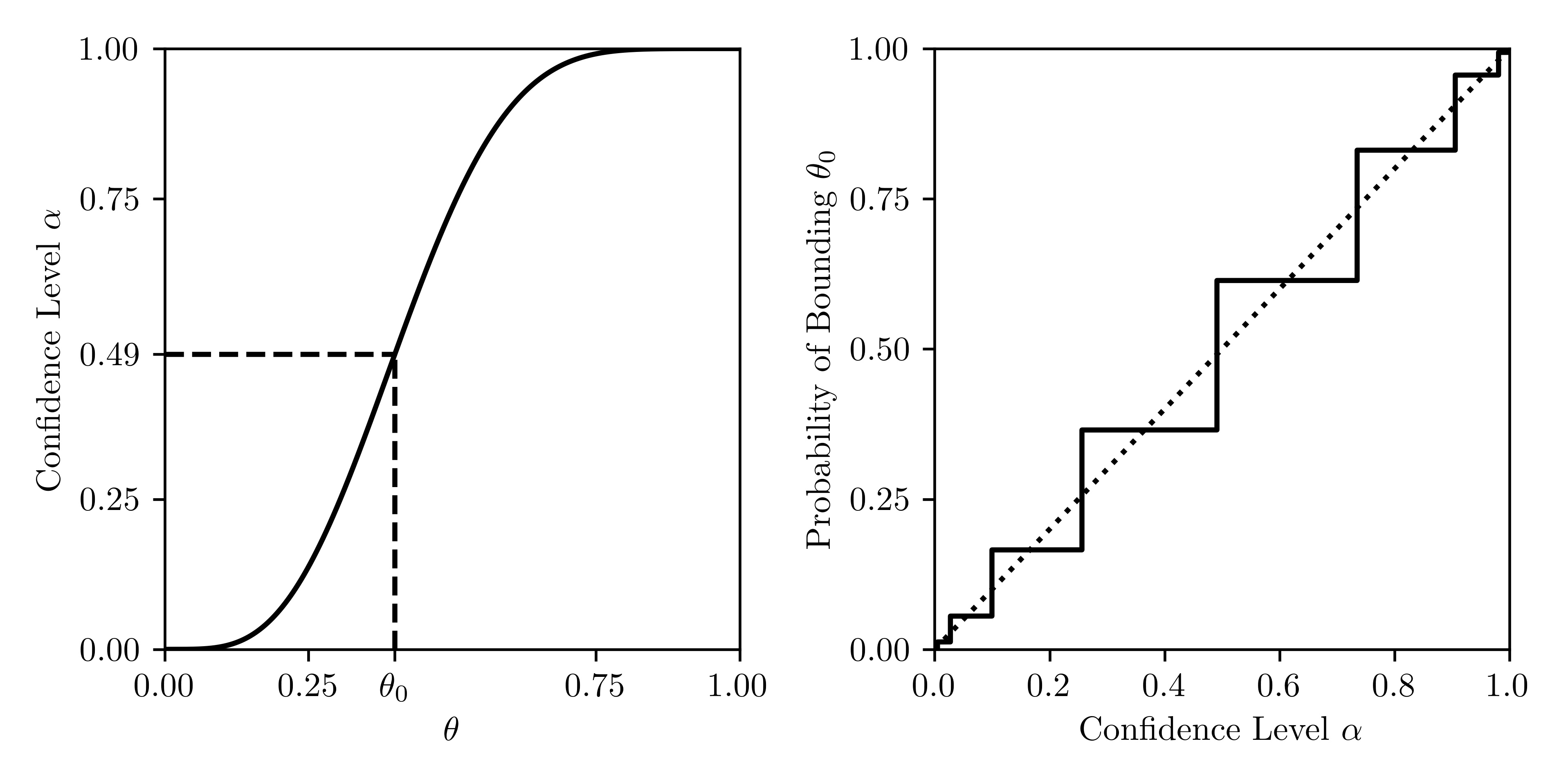}
    \caption{}
    \label{fig:BernCDist}
\end{figure}

Fig. \ref{fig:BernCDist} indicates that the proposed confidence distribution in Equation \ref{eq:BernPPropCDist} is not valid. This is due to the clear deviations from the minimum bounding probability required to maintain coverage, as seen by the Singh plot extending below the U$(0,1)$ plot at many points. This indicates that, for example, an $\alpha=0.45$ level confidence interval would only bound the true value with a minimum rate of $\approx0.33$. This indicates that this structure will often produce intervals which do not bound the true value at the desired rate. In this case, a different structure should be devised.

\subsection{Imprecise Bernoulli Confidence Distribution Singh Plot}\label{sec:ImpConf}
Since Equation \ref{eq:BernPPropCDist} failed to provide coverage, a different structure must be devised. An important aspect of this problem for the case of the Bernoulli distribution with a small sample size of $n=10$ is simply that there is not enough data to make statements of precise confidence levels. A large degree of uncertainty is being ignored in attempting to do so. For the Bernoulli distribution, or its binomial extension, Clopper-Pearson intervals allow each $\theta$ to be assigned an interval of values $\text{C}^*(\theta,\bm{x})=\left[\underline{\alpha},\overline{\alpha}\right]$. However, it should be noted that $\underline{\alpha}$ in this case will be greater than $\overline{\alpha}$, and as such a confidence level cannot be attributed to this interval. Similarly, the inverse operation $\text{C}^{-1*}(\alpha, \bm{x})=[\underline{\theta}, \overline{\theta}]$ returns an interval. This is referred to as an imprecise confidence distribution, or c-box. 

The upper and lower confidence limit distributions $\text{C}^*_U(\cdot)$ and $\text{C}^*_L(\cdot)$ of the c-box define $\underline{\alpha}$ and $\overline{\alpha}$ for each $\theta$. In the case of the Clopper-Pearson c-box, these distributions are defined as follows:
\begin{subequations}\label{eq:ClopPearBounds}
\begin{align}
    \underline{\alpha} & = \text{C}^*_L(\theta,\bm{x}) = \text{B}\left(\theta;\sum{\bm{x}} + 1,n - \sum{\bm{x}}\right)\\
    \overline{\alpha} & = \text{C}^*_U(\theta,\bm{x}) = \text{B}\left(\theta;\sum{\bm{x}}, n - \sum{\bm{x}} + 1\right)
\end{align}

\end{subequations}

Mapping a confidence interval is then performed in a similar manner as for confidence distributions, defining an interval $\left[\underline{\alpha},\overline{\alpha}\right]$ such that $\overline{\alpha}-\underline{\alpha}=\alpha$ and constructing a corresponding interval on $\Theta$ from the minimum and maximum of these intervals:

\begin{subequations}
    \begin{align}
        \text{C}^{*-1}([\underline{\alpha},\overline{\alpha}],\bm{x})&=[\min(\text{C}^{-1}(\underline{\alpha},\bm{x})),\max(\text{C}^{-1}(\overline{\alpha},\bm{x}))]\\
        &=[\text{C}^{*-1}_U(\underline{\alpha},\bm{x}),\text{C}^{-1}_L(\overline{\alpha},\bm{x})]
    \end{align}
\end{subequations}
Naturally, a one-sided $\alpha$ level confidence interval $\left[\underline{\alpha}=0,\overline{\alpha}=\alpha\right]$ produces a confidence interval extending to the minimum of $\theta$, $\text{C}^{*-1}([0,\alpha],\bm{x})=\left[0,\text{C}^{-1}_L(\alpha,\bm{x})\right]$. This can be used as before to construct a Singh plot representing the ability of the lower limit of the c-box to bound the true value. The upper limit is then used to generate the opposite one-sided interval $\left[\underline{\alpha}=1-\alpha,\overline{\alpha}=1\right]$ to produce confidence intervals of $\text{C}^{*-1}([1-\alpha;1],\bm{x})=[\text{C}^{-1}_U(1-\alpha,\bm{x}),1]$.  Again, these values are then ordered and plotted against the unit uniform. Visually this can become cluttered and unappealing as, so an alternative is to use the same lower bounded one-sided $\left[\underline{\alpha}=0,\overline{\alpha}=\alpha\right]$ interval, treating the upper limit of the c-box as an isolated distribution. In this case, the corresponding Singh plot should indicate a total lack of coverage (i.e. all below the unit uniform), since the interval being utilised is effectively the complement of the actual target interval.

Since the Singh plots for the upper and lower limit distributions of the c-box straddle, but never cross, the U$(0,1)$ diagonal the distribution can provide confidence since the true confidence distribution must lie between these bounds. This comes at the cost of wider intervals than the distribution proposed in Equation \ref{eq:BernPPropCDist}. The width of the output intervals will decrease as more information is available. This demonstrates how both precise and imprecise confidence distributions can be developed and assessed for inference on known distributions. However, in cases where distributional assumptions are unjustified, what can be done for non-parametric confidence distributions?

\subsection{Predictive Confidence Distributions}
Singh plots can be used to assess the properties of any confidence structure. The most common examples involve inference, capturing epistemic uncertainty. But the same procedure can be applied to analysis of predictive structures as well. These function similarly to standard confidence distributions used for inference, but they output intervals guaranteed to bound the next drawn sample with a desired frequency rather than bounding some true parameter. There are a number of possible examples, but an interesting imprecise confidence distribution is for non-parametric prediction of the next drawn sample $x_{n+1}$ given a dataset $\bm{x}=\{x_1,\dots,x_n\}$, assuming a continuous distribution.

The procedure for prediction is the same as for inference, though the confidence value of the true subsequent sample $\text{C}(x_{n+1},\bm{x})$ is calculated rather than any fixed parameter. The lower and upper limit confidence distributions are defined as follows for the non-parametric prediction case:
\begin{subequations}\label{eq:EmpDist}
\begin{align}
    \text{C}^*_L(x_{n+1},\bm{x})&=\frac{|\{x_i\in\bm{x}:x_i\leq x_{n+1}\}|}{n+1}\\
    \text{C}^*_U(x_{n+1},\bm{x})&=1-\frac{|\{x_i\in\bm{x}:x_i\geq x_{n+1}\}|}{n+1}
\end{align}
\end{subequations} 

Again, the coverage properties of this structure can be demonstrated using a Singh plot. A gaussian mixture distribution is used here for demonstrative purposes.

This demonstrates that such a structure is at capable of reliably calculating intervals which will contain subsequent samples, at least for this Gaussian mixture model.

\section{Representation of Confidence Structure Characteristics}
The intent of a Singh plot is to rapidly convey the confidence characteristics of the chosen structure, and has been demonstrated that deviations from the central U$(0,1)$ line indicate the coverage probability of a structure. Singh plots are also capable of indicating a number of other characteristics which aid in the design and validation of appropriate imprecise confidence structures.
\subsection{Representing Uncertainty from Limited Data}
\begin{figure}[!ht]
    \centering
    \includegraphics[width=\linewidth]{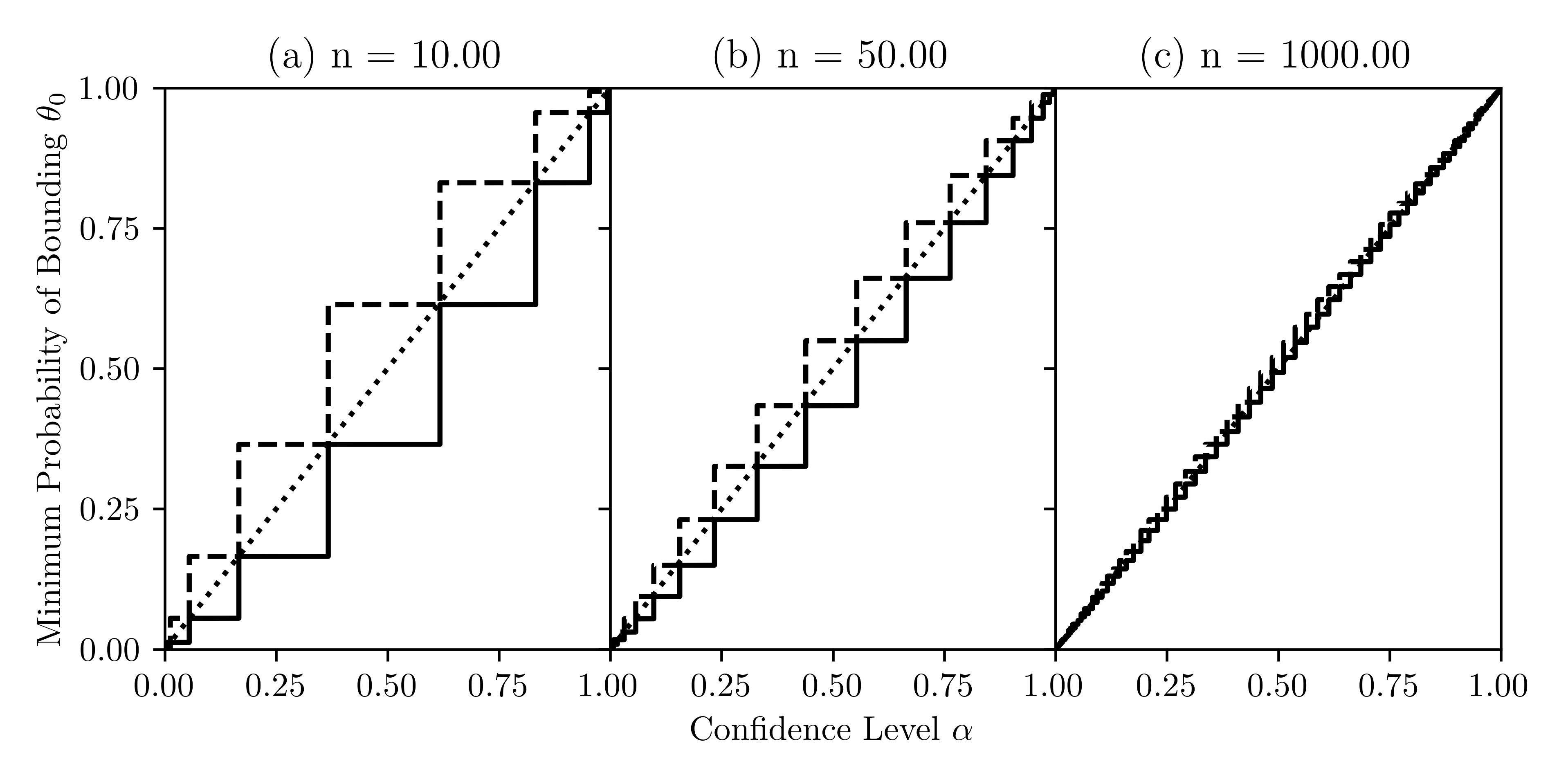}
    \caption{}
    \label{fig:BernSinghComp}
\end{figure}

Figure \ref{fig:BernSinghComp} demonstrates how Singh plots differ when performed on various sample sizes. As the number of data points increases, the imprecise distribution converges to the perfect case, matching the U$(0,1)$ diagonal. Lower samples sizes are shown to produce confidence intervals which have coverage, but which are wider than they would be in the perfect case. For example, with a sample size of $n=10$, an $\alpha=0.65$ confidence interval would have similar coverage properties to one with $\alpha=0.8$.

\subsection{Representing uncertainty about Rare Events}
\begin{figure}[!ht]
    \centering
    \includegraphics[width=\linewidth]{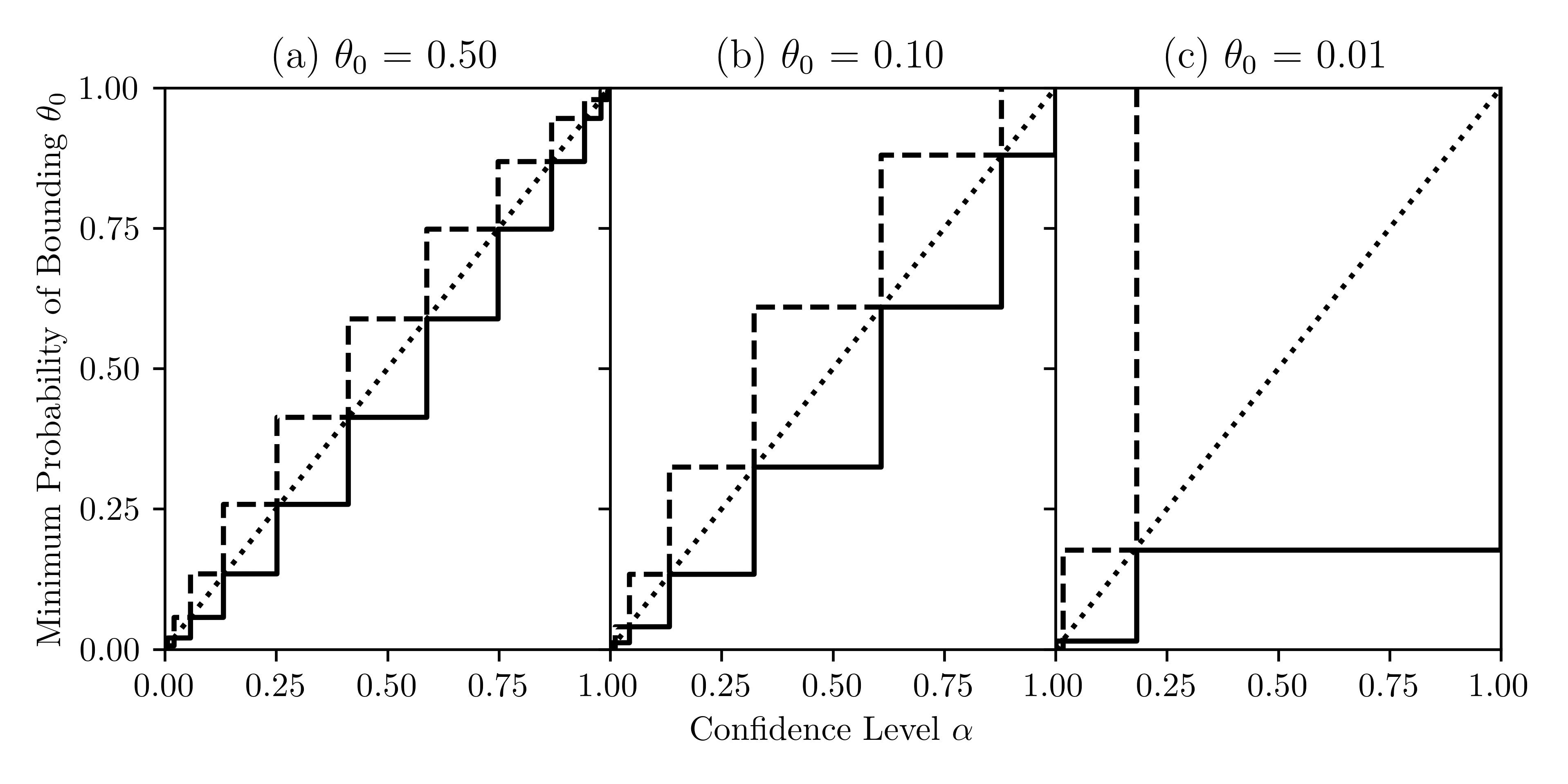}
    \caption{}
    \label{fig:BernSinghCompTheta}
\end{figure}

Figure \ref{fig:BernSinghCompTheta} demonstrates how Singh plots of Equation \ref{eq:BernPPropCDist} respond to varying rate $\theta_0$. Estimation of a very low rate will naturally be difficult when sample sizes are low, and this is reflected in the broad confidence regions for a given bounding probability. For example, an $\alpha=0.2$ confidence interval is as likely to bound the true parameter $\theta_0=0.01$ as one with $\alpha=0.9$. Increasing the sample size will converge these towards the U$(0,1)$ distribution as seen above, though a feature of note is that the Singh plot becomes asymmetrical as $\theta_0$ deviates from the centre of the support $\theta=0.5$.

\subsection{Favourability, Conservatism and Overconfidence}
Since it is relatively simple to produce a Singh plot, they can be used to modify and assess confidence distributions. One hope may be that through some modification, a confidence distribution may be developed which produces tighter bounds whilst preserving the property of coverage. For example, Equation \ref{eq:ClopPearBounds} could be modified to alter the uncertainty expressed in the imprecise confidence distribution. This could be done by using a modification such as that shown in Equation \ref{eq:BopPearBounds} for a length-$n$ sample set $\bm{x}=[x_1,\dots,x_n]$, replacing $c$ with the desired parameter for imprecision.

\begin{subequations}\label{eq:BopPearBounds}
\begin{align}
    \alpha_L & = \text{C}^*_L(\theta,\bm{x}) = \text{B}\left(\theta;\sum{\bm{x}} + c,n - \sum{\bm{x}}\right)\\
    \alpha_U & = \text{C}^*_U(\theta,\bm{x}) = \text{B}\left(\theta;\sum{\bm{x}}, n - \sum{\bm{x}} + c\right)
\end{align}

\end{subequations}
This again produces proposed confidence distributions, as the impact of this change is not yet known. The coverage impact of varying B can then be inspected through the use of Singh plots.

\begin{figure}[!ht]
    \centering
    \includegraphics[width=\linewidth]{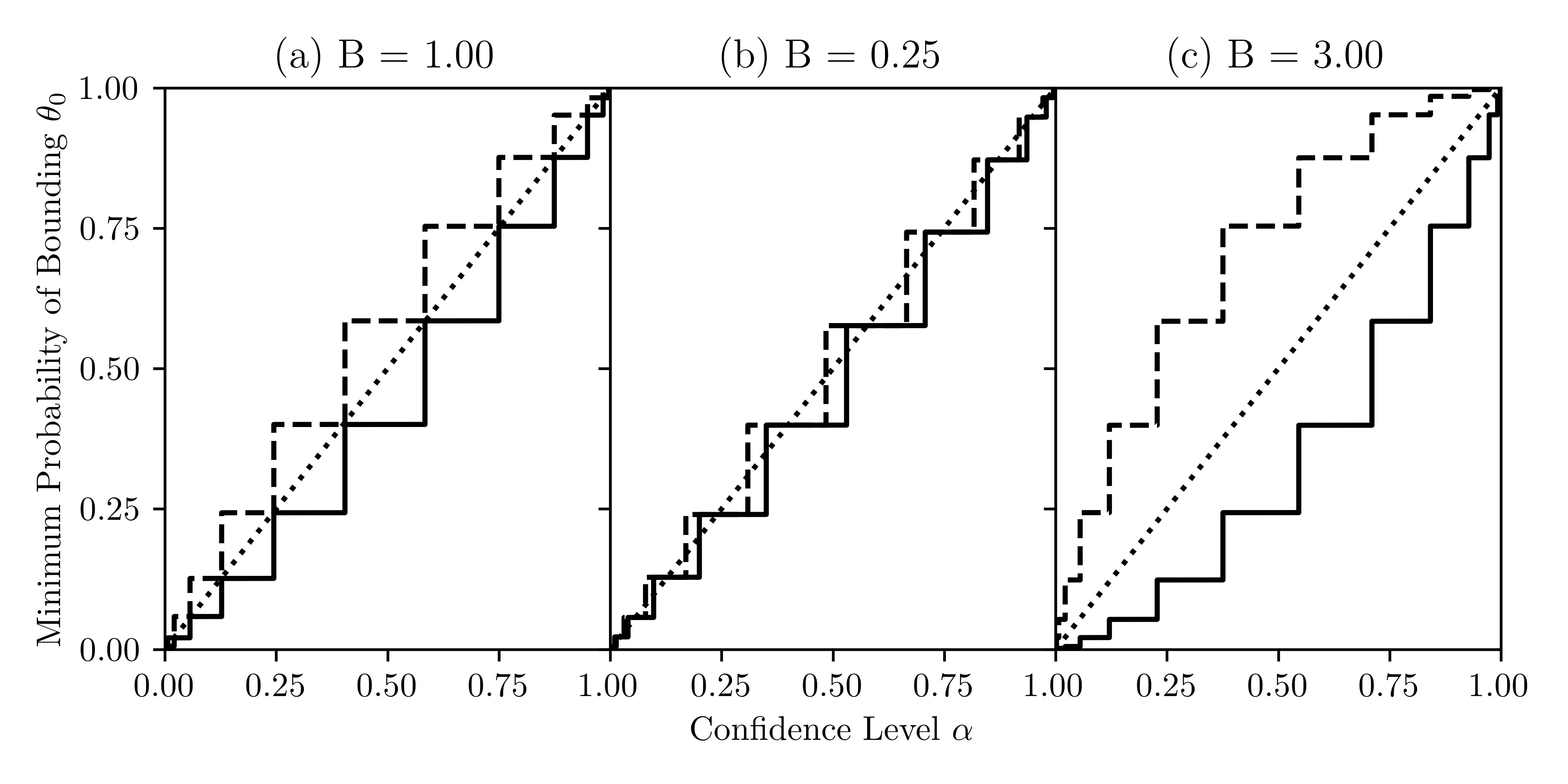}
    \caption{}
    \label{fig:ConfComparison}
\end{figure}

As can be seen in Figure \ref{fig:ConfComparison}, decreasing $c$ beyond 1 produces an invalid c-box, as the Singh plots for both bounds clearly extend beyond the U$(0,1)$ diagonal. This structure would be considered overconfident, as it assigns intervals a level of confidence which is not always exceeded by their coverage probability. 

Increasing $B$ does not produce an invalid structure, but instead produces a structure with additional imprecision. This structure would be considered conservative, as it is assigning intervals a level of confidence which is always greater than but never equal to their coverage probability. A structure such as the case when $c=1$ is still considered conservative by many\cite{balch2020new,singh2007confidence}, since the width of the confidence intervals it produces is wide by comparison to many alternatives. Determining whether a structure is more or less conservative than another in this sense with a Singh plot requires further investigation, though a comparison of the area between the Singh plot bounds should allow for comparisons of relative confidence.

A favourable confidence distribution should coincide with the U$(0,1)$, indicating that further reduction in the width of confidence intervals produced by the chosen distribution has the potential to violate the coverage requirement. This indicates that the confidence distribution is appropriately representing the uncertainty about the parameter of interest. A conservative structure is still suitable for inference, it just implies that the uncertainty about these inferences could be reduced with a more appropriate confidence distribution. An over-confident structure however, cannot be relied upon to produce confidence intervals with coverage, and implies that the applied confidence distribution is neglecting uncertainty.

\begin{figure}[!ht]
    \centering
    \includegraphics[width=\linewidth]{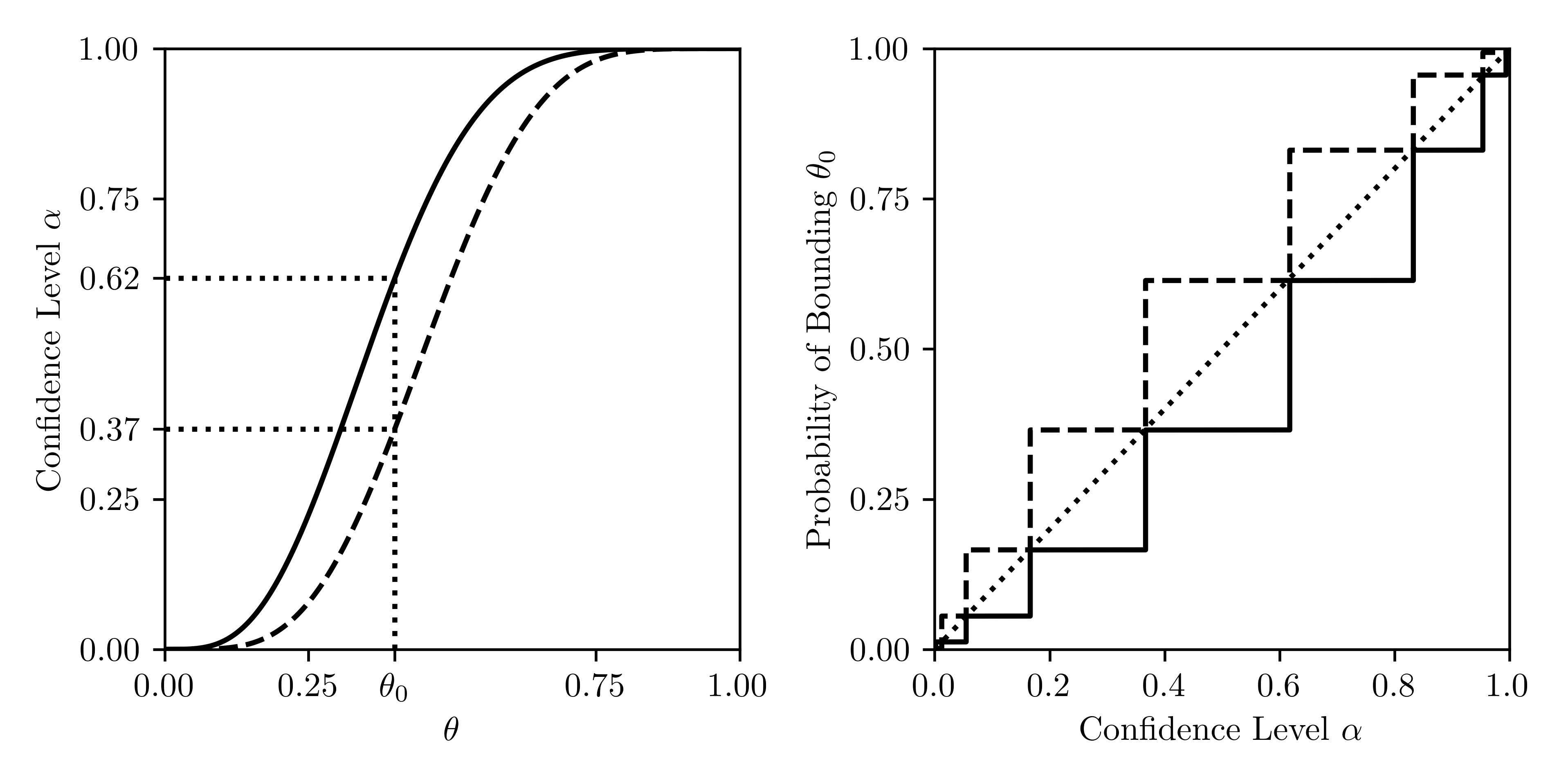}
    \caption{}
    \label{fig:ClopPearDist}
\end{figure}

For a precise distribution, a favourable structure simply converges to U$(0,1)$ as seen in Figure \ref{fig:NormCDist}. For an imprecise structure, this would be instead represented by coincidence with U$(0,1)$ where a step in coverage probability occurs, as seen in Figure \ref{fig:ClopPearDist}.

\subsection{Global Coverage Properties}\label{sec:Global}
Portraying the properties of the confidence box for a known $\theta_0$ allows for insight into the suitability of the chosen structure, but generally $\theta_0$ is unknown, hence the desire to infer its true value. In this case parametric inference implies knowledge of the target distribution, so it is possible to assess the chosen structure across a range of possible representations of the target distribution. In this case, a Singh plot can also portray the global coverage properties for unknown parameters within the support of the parameter $\Theta$. 

This is done by targeting an interval of interest on the support $\bm{\theta}$, sampling this region to generate a series of distributions $\bm{F}=\{\text{F}(\theta_1),\dots,\text{F}(\theta_n)\}$ where each $\text{F}(\theta_i)$ represents the cumulative distribution of the target distribution with parameter $\theta_i$. Samples are generated for each of these distributions, individual Singh plots are calculated and then the lower bound of this second-order Singh plot is used as the final output. If this lower bound satisfies the criteria outlined above for a confidence distribution then the can be used for the target interval with confidence that it will provide coverage, though knowledge of conservatism is lost. If coverage is demonstrated in this case, then the structure can be safely used for any potential case of inference in the interval of interest about the target distribution.

For a precise distribution this can be calculated as follows:

\begin{equation}
    \text{S}_G(\bm{\alpha},\bm{\theta}) = \min_{\theta_i\in\bm{\theta}}\{\text{Pro}(C^{*}(\bm{\theta},\bm{x}\sim\text{F}(\theta_i))\ni\theta_i)\}
\end{equation}

and for an imprecise distribution, where again the lower limit of the Singh plot is inverted for ease of interpretation:

\begin{subequations}
\begin{align}
    \text{S}_L(\bm{\alpha},\bm{\theta}) &= \max_{\theta_i\in\bm{\theta}}\{\text{Pro}(C_L^{*-1}(\bm{\alpha},\bm{x}\sim\text{F}(\theta_i))\ni\theta_i)\}\\
    \text{S}_U(\bm{\alpha},\bm{\theta}) &= \min_{\theta_i\in\bm{\theta}}\{\text{Pro}(C_U^{*-1}(\bm{\alpha},\bm{x}\sim\text{F}(\theta_i))\ni\theta_i)\}
\end{align}
    
\end{subequations}

\begin{figure}[!ht]
    \centering
    \includegraphics[width=\linewidth]{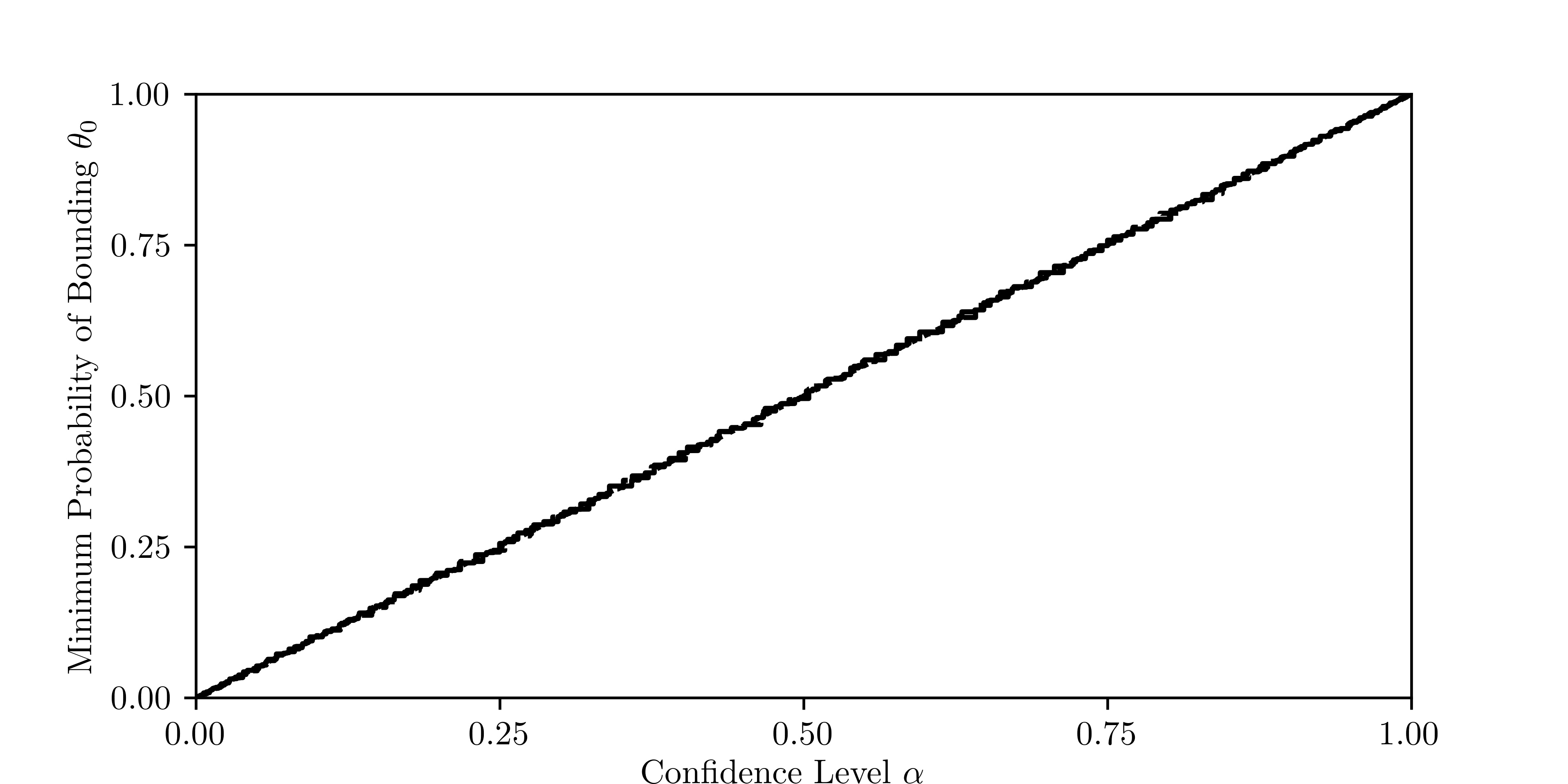}
    \caption{}
    \label{fig:GlobSingh}
\end{figure}

This is demonstrated in Figure \ref{fig:GlobSingh} with the structure described in Equation \ref{eq:ClopPearBounds}, taking values of $\theta$ across the interval [0,1]. This demonstrates that the Clopper-Pearson confidence structure is capable of providing intervals with frequentist coverage regardless of the value of $\theta$. An example of how to perform this is given in Algorithm \ref{algo:global}.

\IncMargin{1em}
\begin{algorithm}
    \SetAlgoLined
    \SetKwInOut{Input}{input}\SetKwInOut{Output}{output}
    \Input{$C^*\leftarrow$Proposed confidence structure\\
        $\text{f}(\bm{\theta})\leftarrow$Target distribution taking parameters $\theta$\\
        $\{\theta_0, \dots, \theta_k\}\leftarrow$Parameter sets of interest\\
        $\theta_{l,0}\leftarrow$True values for parameter of interest in set $l$\\
        $\theta_{l,1:k}\leftarrow$True values for nuisance parameters in set $l$}
    \Output{Singh plot for visual assessment of global confidence structure properties}
    \For{$j\in1,\dots,k$}{
    \For{$i \in {1,\dots,m}$}{
        Generate sample: $\bm{x}={x_1, \dots, x_n}\sim\text{f}(\theta_{j})$\\
        Calculate minimum required confidence for coverage: $t_{j,i} = \text{C}^*(\theta_{i,0},\bm{x})$\\
    }
    Sort $t_{j,1:m}$\\
    }
    \For{$i \in {1,\dots,m}$}{
        Estimate global minimum required confidence for coverage: $s_i=\min(t_{1:k,i})$\\
    }
    Plot empirical CDF of $s$\\
    Plot CDF of $U(0,1)$ for comparison
    \caption{Generation of a global Singh plot}\label{algo:global}
\end{algorithm}\DecMargin{1em}

It should be noted that the sample size will also affect the inference about $\theta_0$. It is assumed that this structure is applied to a case with a known sample size, otherwise an exhaustive representation of the minimum coverage probability would have to be calculated by sampling combinations of $(\theta_0, n)$. Similarly, nuisance parameters may be treated in a similar manner, though the most efficient means of doing so will vary depending on the structure being assessed. Calculating Singh plots with variations in the nuisance parameters may indicate whether their effect on the minimum confidence required is monotone, and if so end-points could be taken to reduce computational cost.

\section{Chebyshev UCL Coverage}
Most of the above example are demonstrations of known confidence structures or of clearly deficient suggestions. However, the value of Singh plots lies in visual demonstrations of the performance of structures where the deficiencies are not known.

As an example, the ProUCL package is a software package for statistical analysis of environmental datasets, and one of the statistics that can be calculated is the upper confidence limit of the mean of a population, $\bar{\mu}\in M$. This could be used for calculation of the upper confidence limit on the expected value of the concentration of a particular pollutant in water samples, amongst many other use cases. The software documentation notes the difficulty of handling skewed datasets and suggests the use of an estimator based on the Chebyshev inequality, defined below in Equation \ref{eq:ProUCLCheb}\cite{Asinghs}.

\begin{equation}\label{eq:ProUCLCheb}
    \bar{\text{C}^*}(\bm{\alpha}, \bm{x}) = \mu_{\bm{x}} + \sqrt{\frac{1}{1-\alpha}-1}\frac{\sigma_{\bm{x}}}{\sqrt{n}}
\end{equation}

The novelty of this upper confidence limit is the claim that it is a reasonable non-parametric estimator, that is it should be correct regardless of the underlying distribution. This is an excellent quality for an estimator to have, though the documentation of ProUCL does note that highly skewed datasets may lose coverage, and that in such cases the data should be inspected to ensure that there is truly only a single population being reported. This raises questions about how skewness affects the coverage, and is it really reasonable to simply raise or lower the required $\alpha$ level to get an appropriate confidence interval?

Singh plots can serve here as a tool for inspecting the properties of this estimator in an intuitive manner. A family of distributions can be generated to `stress test' the provided estimator. In this case, scaled Bernoulli distributions represent a family of distributions which should be particularly difficult for such an estimator to maintain coverage. The estimator relies on scaling the standard deviation of a sample set, and there are many sample sets that can be drawn with a high probability from a highly skewed Bernoulli process which have zero standard deviation. The Bernoulli parameter $p$ here can be manipulated to alter the skewness of the distribution in order to observe how highly skewed datasets affect the coverage of this confidence limit.

The PRoUCL version 5.1.0 documentation defines 'extremely skewed' as data where the standard deviation of the log transformed data $\hat{\sigma}_{\bm{x}}$ is greater than 3. For a Bernoulli distribution, this statistic is inverse to the observed skewness since the maximum standard deviation will be observed where $p=0.5$ and the distribution has no skewness.

Firstly Equation \ref{eq:ProUCLCheb} must be inverted to map $\mu\in M$ onto the support of a $\bm{\alpha}$. This gives Equation \ref{eq:ProUCLConf}:

\begin{equation}\label{eq:ProUCLConf}
    \bar{\text{C}^*}^{-1}(\mu, \bm{x}) = 1 - \left(\left(\frac{\sqrt{n}}{\sigma_{\bm{x}}}-\mu_{\bm{x}}\right)^2+1\right)^{-1}
\end{equation}

This can then be used to generate a Singh plot for a variety of Bernoulli distributions with skewness controlled by the parameter $p=\theta$. According to the ProUCL documentation, it should be expected that the structure provides coverage for moderately skewed data, but that this may not hold for highly skewed data.

For small sample sizes, Equation \ref{eq:ProUCLCheb} fails to provide coverage at all confidence levels even for the unskewed Bernoulli distribution ($p=0.5$, skewness = 0). Any skew in the dataset detracts further from the ability to provide coverage. This can be offset with a larger sample size, though even with 30 samples skewed data $(p=0.05$, skewness = 4.13) leads to a lack of coverage. As such, Equation \ref{eq:ProUCLCheb} should not be considered for use on small datasets, particularly those which may be skewed. A 95\% upper confidence limit from this structure cannot be guaranteed to bound the true mean at least 95\% of the time. In the $p=0.2$ case (skewness=1.5) with a sample size of $n=5$ for example, a 95\% confidence interval would provide only ~67\% coverage. These coverage figures also only apply to the particular distributions they are applied to. In practice the utility of this estimator comes from it's supposed applicability to non-parametric cases. Because of this, attempting to suggest a lower $\alpha$ level in order to be more accurate to the true coverage, or a higher $\alpha$ level to try and be more conservative would not be justifiable.

This upper confidence limit estimator may be of some use, but it is in no way a distribution-free estimator and should not be used for these purposes when sample sizes are small. However, Singh plots may be a means of determining the limits of its use as a confidence estimator and in this case it appears that increasing the sample size allows for confidence on mildly skewed datasets. Whether a practitioner wants to accept the conservatism and the potential for losing applicability to highly skewed datasets is a matter of choice, but Singh plots such as these may be a useful means of informing this decision.

\section{Conclusion}
Singh plots, whilst not technically capable of providing strict proof of coverage, represent an intuitive and simple means of portraying the coverage properties of confidence structures, both precise and imprecise. They allow for comparisons against different proposed structures, as well as analysis of general and specific cases for inference and prediction.

Confidence structures are a widely applicable means of providing probabilistic statements, and Singh plots allow for their use and development without requiring specialist knowledge regarding their formulation. This allows for more widespread adoption and development of this robust approach to uncertainty quantification. This is particularly relevant for the development of procedures suitable for calculating with confidence structures.

\newpage
\section*{Figures}

\begin{figure}[!ht]
    \centering
    \includegraphics[width=\linewidth]{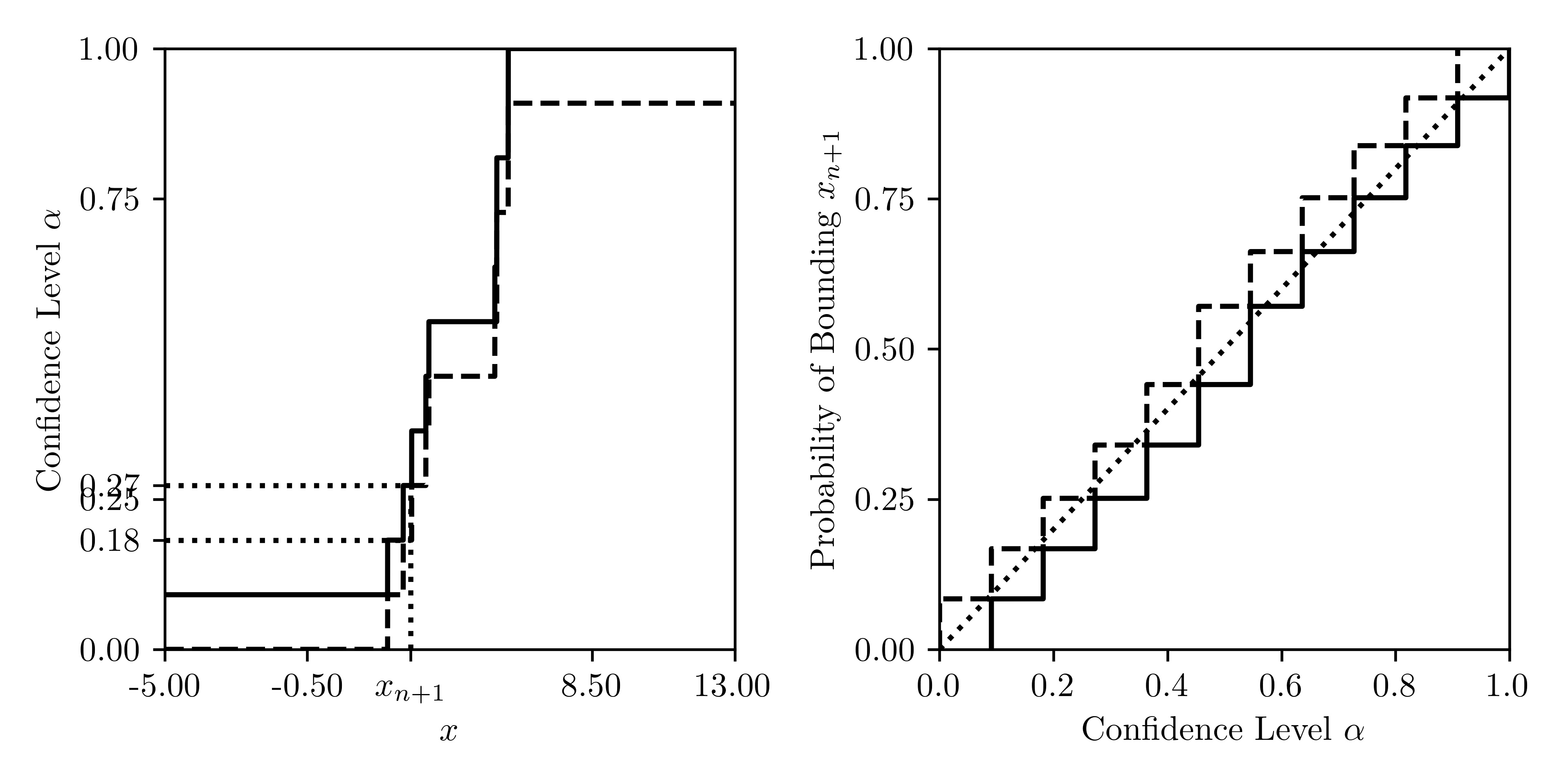}
    \caption{}
    \label{fig:EmpDist}
\end{figure}
\begin{figure}[!ht]
    \centering
    \includegraphics[width=\linewidth]{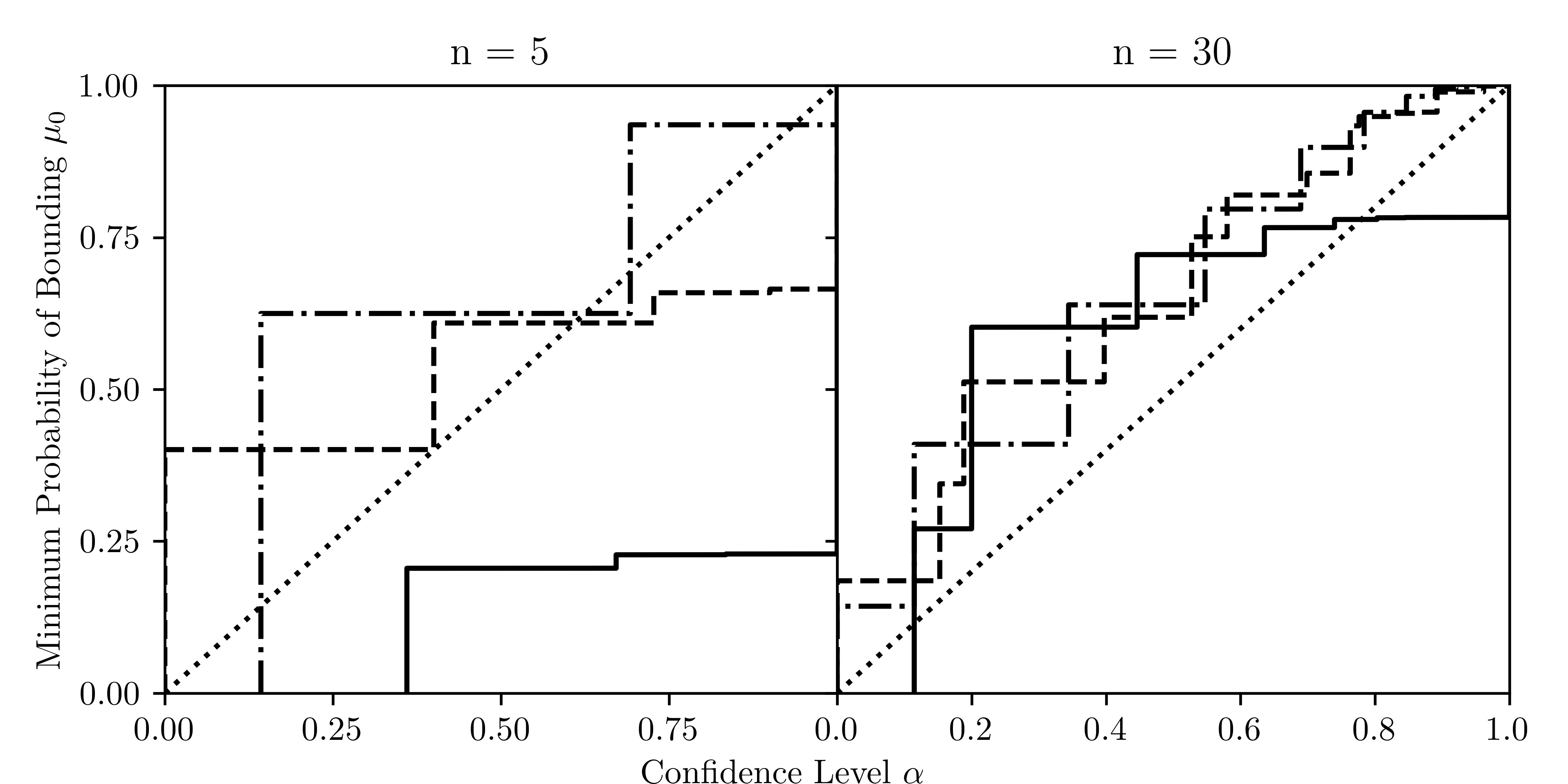}
    \caption{}
    \label{fig:ProUCLDoesntWork}
\end{figure}
\clearpage
\newpage
\section{Figure Captions}
\begin{enumerate}
    \item (a): A single example of a proposed confidence distribution from Equation \ref{eq:NormPivot} generated from $\bm{x}=\{x_1,\dots,x_{10}\}\sim\text{N}(\mu_0=4,\sigma=3)$. The confidence required for a one-sided interval to cover the true mean $\mu_0$ in this example is shown as 0.47. (\textit{solid}, $\text{C}^*(\bm{\mu},\bm{x})$; \textit{dashed}, $\text{C}^*(\mu_0,\bm{x})$). (b): Singh plot for the proposed confidence distribution about the same target distribution, generated from $m=10^4$ sample sets $\bm{X}=\{ \bm{x}_1, \dots, \bm{x}_m\}$. (\textit{solid}, $\text{S}(\bm{\alpha};\mu_0)$; \textit{dashed}, $\text{U}(0,1)$).

    \item (a): A proposed confidence  distribution from Equation \ref{eq:BernPPropCDist} generated from $\bm{x}=\{x_1,\dots,x_{10}\}\sim\text{Bin}(N=1,p=\theta_0)$. The confidence value of the true rate $\theta_0$ is shown as 0.49. (\textit{solid}, $\text{C}^*(\bm{\theta},\bm{x})$; \textit{dotted}, $\text{C}^*(\theta_0,\bm{x})$). (b): Singh plot for the proposed confidence distribution about the same target distribution, generated from $m=10^4$ sample sets $\bm{X}=\{ \bm{x}_1, \dots, \bm{x}_m\}$. (\textit{solid}, $\text{S}(\bm{\alpha};\theta_0)$; \textit{dotted}, $\text{U}(0,1)$).

    \item (a): A proposed confidence  distribution from Equation \ref{eq:ClopPearBounds} generated from $\bm{x}={x_1,\dots,x_{10}}\sim\text{Bin}(n=1,p=\theta_0)$. The confidence value interval of the true rate $\theta_0$ is shown as [0.37, 0.62]. (\textit{solid}, $\text{C}_U^*(\bm{\theta},\bm{x})$;  \textit{dashed}, $\text{C}_L^*(\bm{\theta},\bm{x})$; \textit{dotted}, $\text{C}^*(\theta_0, \bm{x})$). (b): Singh plot for the proposed confidence distribution about the same target distribution, generated from $m=10^4$ sample sets $\bm{X}=\{ \bm{x}_1, \dots, \bm{x}_m\}$. (\textit{solid}, $\text{S}_U(\bm{\alpha};\theta_0)$; \textit{dashed}, $\text{S}_L(\bm{\alpha};\theta_0)$; \textit{dotted}, $\text{U}(0,1)$).
    
    \item (a): A proposed confidence distribution from Equation \ref{eq:EmpDist} generated from a length $n=10$ sample set $\bm{x}=\{x_1,\dots,x_{10}\}\sim \text{F}([\mu_1,\mu_2],[\sigma_1,\sigma_2])$ where $\text{F}([\mu_1,\mu_2],[\sigma_1,\sigma_2])=0.5\cdot\text{N}(\mu_1=4,\sigma_1=3) + 0.5\cdot\text{N}(\mu_2=5,\sigma_2=1.5)$. The confidence value of the true value $x_{n+1}$ is shown as $\text{C}(\mu_0,\bm{x})=[0.18,0.27]$. (\textit{solid}, $\text{C}_U^*(\bm{\theta},\bm{x})$;  \textit{dashed}, $\text{C}_L^*(\bm{\theta},\bm{x})$; \textit{dotted}, $\text{C}^*(\theta_0, \bm{x})$).  (b): Singh plot for the proposed imprecise confidence distribution about the same target distribution, generated from $m=10^4$ sample sets $\bm{X}=\{ \bm{x}_1, \dots, \bm{x}_m\}$. (\textit{solid}, $\text{S}_U(\bm{\alpha};\theta_0)$; \textit{dashed}, $\text{S}_L(\bm{\alpha};\theta_0)$; \textit{dotted}, $\text{U}(0,1)$).
    
    \item A series of Singh plots used for inference about $\theta_0=0.4$ generated using Equation \ref{eq:BernPPropCDist} and a dataset of varying length $n$. (\textit{solid}, $\text{S}_U(\bm{\alpha};\theta_0)$; \textit{dashed}, $\text{S}_L(\bm{\alpha};\theta_0)$; \textit{dotted}, $\text{U}(0,1)$).
    
    \item A series of Singh plots used for inference about a varying $\theta_0$ generated using Equation \ref{eq:BernPPropCDist} and a length $n=20$ dataset. (\textit{solid}, $\text{S}_U(\bm{\alpha};\theta_0)$; \textit{dashed}, $\text{S}_L(\bm{\alpha};\theta_0)$; \textit{dotted}, $\text{U}(0,1)$).
    
    \item A series of Singh plots used for inference about $\theta_0=0.4$ generated using Equation \ref{eq:BernPPropCDist} and a length $n=20$ dataset with varying degrees of confidence demonstrated by altering the $c$ parameter in Equation \ref{eq:BopPearBounds}. (\textit{solid}, $\text{S}_U(\bm{\alpha};\theta_0)$; \textit{dashed}, $\text{S}_L(\bm{\alpha};\theta_0)$; \textit{dotted}, $\text{U}(0,1)$).
    
    \item Global Singh plot produced using $m=100$ $\theta$ samples drawn from $[0,1]$ and $N=10^3$ Monte Carlo samples using the Clopper-Pearson confidence structure for inference about $\theta$ with a sample size of 10. (\textit{solid}, $\text{S}_U(\bm{\alpha};\theta_0)$; \textit{dashed}, $\text{S}_L(\bm{\alpha};\theta_0)$; \textit{dotted}, $\text{U}(0,1)$).
    
    \item Singh plots representing the coverage probability for a desired $\alpha$ confidence level interval using Equation \ref{eq:ProUCLConf} for inference about data generated from Bernoulli distributions with varying $p$-parameters, scaled to have a consistent mean of $\mu_0=2$ and a minimum value of $\min{M}=0$. Two plots are shown, for sample sizes of n=5 (left) and n=30 (right). (\textit{solid}, $\text{S}(\bm{\alpha};p_0=0.05)$; \textit{dashed}, $\text{S}(\bm{\alpha};p_0=0.2)$; \textit{dash-dot}, $\text{S}(\bm{\alpha};p_0=0.5)$; \textit{dotted}, $\text{U}(0,1)$).
\end{enumerate}
\end{document}